# Strong exciton−photon coupling with colloidal quantum dots in a tuneable microcavity

Dmitriy Dovzhenko,[1,2, a)] Maksim Lednev,[1] Konstantin Mochalov,[3] Ivan Vaskan,[3,4] Pavel Samokhvalov,[1] Yury Rakovich,[5] and Igor Nabiev[1,6,a)*]

[1] *National Research Nuclear University MEPhI (Moscow Engineering Physics Institute) 115409 Moscow, Russian Federation*

[2] *Department of Physics and Astronomy, University of Southampton, Southampton, SO17 1BJ, United Kingdom*

[3] *Shemyakin–Ovchinnikov Institute of Bioorganic Chemistry, Russian Academy of Sciences, 117997 Moscow, Russian Federation*

[4] *Moscow Institute of Physics and Technology, Dolgoprudny, 141701 Moscow, Russian Federation*

[5] *IKERBASQUE, Basque Foundation for Science, 48009 Bilbao; Donostia International Physics Center; Polímeros y Materiales Avanzados: Física, Química y Tecnología, UPV-EHU; Centro de Física de Materiales (MPC, CSIC-UPV/EHU) 20018 Donostia - San Sebastian, Spain*

[6] *Laboratoire de Recherche en Nanosciences, LRN-EA4682, Université de Reims Champagne-Ardenne, 51100 Reims, France*

[a)] Authors to whom correspondence should be addressed: dovzhenkods@gmail.com or igor.nabiev@univ-reims.fr

**ABSTRACT**

Polariton emission from optical cavities integrated with various luminophores has been extensively studied recently due to the wide variety of possible applications in photonics, particularly promising in terms of fabrication of low-threshold sources of coherent emission. Tuneable microcavities allow extensive investigation of the photophysical properties of matter placed inside the cavity by deterministically changing the coupling strength and controllable switching from weak to strong and ultra-strong coupling regimes. Here we demonstrate room-temperature strong coupling of exciton transitions in CdSe/ZnS/CdS/ZnS colloidal quantum dots with the optical modes of a tuneable low-mode-volume microcavity. Strong coupling is evidenced by a large Rabi splitting of the photoluminescence spectra depending on the detuning of the microcavity. A coupling strength of 154 meV has been achieved. High quantum yields, excellent photostability, and scalability of fabrication of QDs paves the way to practical applications of coupled systems based on colloidal QDs in photonics, optoelectronics, and sensing.

Reversible coherent energy exchange between the exciton transition and electromagnetic field modes in an optical cavity leads to the formation of quasi-particles called polaritons when the rate of the energy exchange exceeds the rate of losses in the coupled system.[1] This exciton–photon hybridization of the states within the cavity can significantly alter their intrinsic properties of a quantum emitters, including energy,[2] lifetime,[3] emission spectra, and the possibility of energy transfer.[4,5] The bosonic nature of polaritons together with their extremely low effective mass have been shown to promote the formation of non-equilibrium condensates.[6] Superfluidity of the polariton condensate has also been demonstrated.[7] Since the first demonstration of polariton lasing, many systems have been shown to be suitable for operation in the strong coupling regime and to form polariton condensates. Further improvement has been achieved with the use of organic materials. Frenkel excitons in these materials, in contrast to the Wannier–Mott excitons in inorganic systems, having considerably higher binding energies, allow room-temperature operation of strongly coupled systems.[8-10] Semiconductor nanoparticles represent a unique example of inorganic materials that can be used to achieve strong coupling at room temperature. Flatten et. al.[11] were the first to demonstrate strong exciton–photon coupling using colloidal semiconductor nanoplatelets. The achieved value of Rabi splitting was as large as 66 and 58 meV for heavy and light holes, respectively. Zero-dimensional colloidal nanocrystals, or quantum dots (QDs), although being characterized by lower oscillator strengths,[12,13] have the advantage of high fluorescence quantum yields and small physical volume,[14] causing a plethora of their practical applications in sensing, optoelectronics and nanomedicine.[15] Therefore, QDs are promising inorganic materials for obtaining room-temperature strong coupling and polariton condensation. Due to a high quantum yield, relatively large transition dipole moment (up to 100-200 D),[16,17] high photostability, and well-developed technology of their large-scale fabrication, QDs could be used in various polariton-based devices. We have used a tuneable microcavity to investigate the formation of strongly coupled states in an ensemble of QDs placed between two metal mirrors. The tuneable microcavity was developed previously and has already demonstrated the capacity for achieving high values of coupling strength.[18] The lateral localization of the optical mode and several orders of magnitude lower values of mode volume has been achieved with the use of an upper mirror with a convex surface.[19,20]

In this study, we have obtained strong coupling between exciton transition in colloidal QDs and the optical modes of the tuneable microcavity. We have measured the dependence of the QD PL spectrum on the separation distance between the mirrors of the tuneable microcavity, where emission from the weakly coupled part of the QD ensemble was observed along with the emission from the lower polariton branch. We have analysed the polariton dispersion and extracted the value of the coupling strength of about 154 meV. Furthermore, we have measured the dependence of the emission on the power of non-resonant excitation. At the same time, we have observed neither narrowing of the lower polariton emission peak nor threshold behaviour. Instead, saturation of the emission has been demonstrated as the pump power was increased.

The tuneable microcavity, used in our experiments, has been described in detail previously.[18,20] Briefly, it consists of two metal mirrors with a reflection of about 87%: convex (curvature radius, 77.3 mm) upper and a flat bottom mirrors, the latter positioned on a piezo stack. The thickness of the aluminium layer of the mirrors was about 35 nm, with a 20-nm protective SiO2 layer. This configuration forms an unstable Fabry–Perot microcavity with relatively low Q-factors. The advantage of this approach is that the mode volume can be as small as several units of $\left(\frac{\lambda}{n}\right)^3$, where $\lambda$ is the mode wavelength and $n$ is the refractive index of the media inside the cavity.[19]

The rough positioning in all three directions was performed by means of several micrometre screws, and the fine positioning with a precision of several nanometres was done by changing the voltage applied to the piezo stack while simultaneously controlling the transmission spectrum. This configuration allowed us to establish the laterally confined optical mode in the area of the lowest point of the upper convex mirror with a relatively low mode volume. The details of the experimental setup consisting of a tuneable microcavity could be found elsewhere.[18-20] The tuneable microcavity as a part of the unique setup called the *System for Probe-Optical 3D Correlative Microscopy* (http://ckp-rf.ru/usu/486825/) was mounted on an inverted Ntegra-base (NT-MDT) confocal microspectrometer with a 100X/0.80 MPLAPON

XYZ piezo positioned lens (Olympus). An MCWHF2 white LED (Thorlabs) with an optical condenser was used for transmission measurements. For non-resonant PL excitation, we used emission of a 450-nm CW laser (L450P1600MM, Thorlabs). The recording system included an Andor Shamrock 750 monochromator equipped with an Andor DU971P-BV EMCCD (Andor Technology Ltd.). Typical transmission spectrum of the microcavity is shown in Fig. 1a. The Q-factor of the mode was about 140, with the mode volumes estimated to be in the range from 15 to 20 units of $\left(\frac{\lambda}{n}\right)^3$.

The space between the microcavity mirrors was filled with immersion oil solution containing semiconductor CdSe(core)/ZnS/CdS/ZnS(multishell) QDs at a concentration of about 0.52 mg/ml. QDs were fabricated using the hot injection method.[14] For preparation of a homogeneous solution of QDs in immersion oil, 50 µL of a QD solution in hexane was mixed with the same volume of immersion oil, sonicated for 5 min, and put into a water bath at 80°C for 10 min in order to evaporate hexane from the mixed solutions. After the hexane evaporation, the QD solution in immersion oil was once again sonicated in order to achieve homogeneity. For experiments, 10 µL of each sample was placed on the flat bottom mirror of the microcavity and then covered with the convex upper mirror. During the experiments, the sample solvent evaporation was negligible, and the sample concentration was constant. The absorption and photoluminescence spectra of the QD solution outside of the cavity, as well as their TEM image, are shown in Fig. 1.

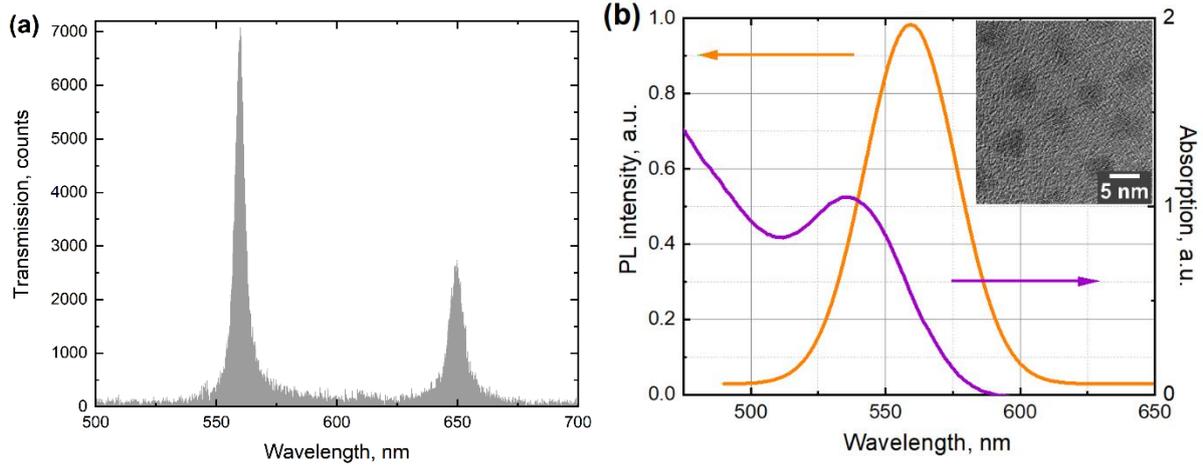

**FIG. 1.** Tuneable microcavity transmission spectra (a); the absorption (purple) and photoluminescence (orange) spectra of CdSe(core)/ZnS/CdS/ZnS(multishell) QDs in a solution, with a TEM image of QDs in the inset (b).

The absorption spectrum of the QDs consists of a wide absorption band in the short-wavelength region and the first exciton peak at about 535 nm (2.32 eV). The PL spectrum of the QDs represents a symmetrical Gaussian curve with a maximum at 560 nm (2.21 eV) and a full width at half maximum (FWHM) of about 40 nm. In principle, in an unstable microcavity, used in this study, additional mode localization in the plane perpendicular to the main axis of the cavity is achieved due to the losses determined by the curved surface of the upper mirror. According to the numerical simulation, the spatial region of mode localization at the point of the minimum distance between the mirrors is several micrometres, which is much smaller than the 300-μm focal spot of the objective lens used for collection of the signal in both transmission and PL experiments. Consequently, in the transmission measurements, most of the light passes through the cavity uncoupled, because of the much larger number of photons entering the cavity as compared to the number of strongly-coupled QDs. On the other hand, in the case of PL measurements, the number of emitting QDs cannot exceed the total number of QDs distributed within the optical mode. Thus, despite the large values of splitting of the PL spectra under non-resonant laser excitation, we should not necessarily expect a parallel observation of splitting in the transmission spectra.

The dependence of the PL spectra of the hybrid system on the cavity mode detuning is shown in Fig. 2(a). The cavity mode tuning has been performed by changing the distance between the microcavity mirrors from 1178 to 1386 nm.

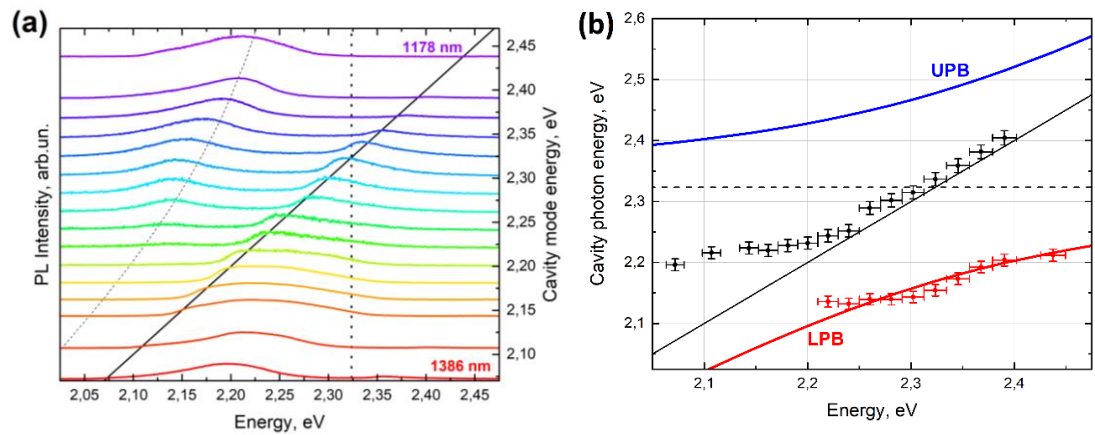

**FIG. 2.** The PL spectra (a) recorded at distances between the mirrors varied from 1178 to 1386 nm (top–bottom) for a microcavity filled with the solution of CdSe(core)/ZnS/CdS/ZnS(multishell) QDs. The black solid line shows the position of the cavity mode; the black dotted line shows the positions of the QD first exciton transition; the black dashed line shows the calculated lower polaritonic branch dispersion. The energies of the lower (red), and upper (dark blue) polariton branches at different cavity detunings (b), experimentally derived from the PL spectra (dots) and theoretically calculated (solid lines). The black dots correspond to the emission from weakly coupled QDs. The horizontal dashed line shows the positions of the QD first exciton transition, the solid black line shows the cavity mode energy.

For the maximum cavity detuning from the QD exciton transition, we have observed an almost unchanged PL spectrum. However, it should be noted that, for the largest distance between the mirrors, when the cavity mode is largely negatively detuned from the QD exciton, a small peak arose at about 2.36 eV. This could be attributed to the weak coupling of the highest energy part of inhomogeneously broadened exciton transitions to the higher-order longitudinal optical mode. It is also important that the free spectral range between this mode and the cavity mode was large enough to neglect the influence of the neighbouring longitudinal modes on the PL spectra of the QDs at all other cavity detunings. When the distance between the mirrors decreased and the cavity photon energy reached the exciton transition energy, two changes occurred. First, the weakly coupled part of the QD ensemble produced emission becomes enhanced at the wavelength corresponding to the microcavity mode and slightly shifted to the blue region due to the presence of higher-order transverse modes in the microcavity. Second, a PL peak corresponding to the lower polariton branch became visible first at 2.13 meV and then shifted towards the bare exciton energy with rising energy of the cavity mode. As the cavity mode shifted to higher energy and the overlap between the bare exciton emission and cavity mode disappeared, the emission from the bare excitonic states merged with the lower polariton branch. Typically, the emission from the upper polariton branch (UPB) in optical microcavities filled with emitters is not observed due to the fast relaxation of the excitation from the UPB to the exciton reservoir.[9] Thus, the lower polariton branch (LPB) energy displays anticrossing with the emission from weakly coupled emitters, which appears at the bare cavity mode energy.

The light–matter interaction observed in the present research was described in terms of a Jaynes–Cummings Hamiltonian in which quantized electromagnetic field is coupled with a two-level system corresponding to the emitter. It can be written in the following way:

$$H = h\nu_{cav} a^\dagger a + \frac{1}{2} h\nu_x \sigma_z + hg(a^\dagger \sigma_- + a\sigma_+), \qquad (1)$$

where $h\nu_{cav}$ and $h\nu_x$ are the energies of the unperturbed cavity mode and emitter absorption transition, respectively; $hg$ is the coupling strength representing the rate of energy exchange between the light and matter constituents; $a(a^\dagger)$ is the annihilation (creation) operator for a cavity photon, $\sigma_{z,-,+}$ are operators acting on the emitter part of the wavefunction; $\sigma_z = |e\rangle\langle e| - |g\rangle\langle g|$, $\sigma_- = |g\rangle\langle e|$, and $\sigma_+ = |e\rangle\langle g|$ ($|g\rangle$ and $|e\rangle$ are the wavefunctions of the ground and excited states of the emitter, respectively). In a single excitation subspace the diagonalization of the Hamiltonian gives two eigenvalues $E_\pm$:

$$E_\pm = \frac{h\nu_{cav} + h\nu_x}{2} \pm \frac{1}{2}\sqrt{(h\nu_{cav} - h\nu_x)^2 + 4(hg)^2}. \qquad (2)$$

To obtain the value of coupling strength, we used the experimental frequencies of the lower polaritonic state as $E_-$ and, by varying the free parameter $g$, found the best fit of our experimental data. This approach was implemented by means of the differential evolution method.[21] The obtained value of coupling constant $g$ is 154 meV. It should be stressed that this value is significantly higher than the coupling strength reported for colloidal nanoplatelets in an open microcavity, despite the considerably higher dipole moment of exciton transition in nanoplatelets. Indeed, previously reported values of coupling strength for semiconductor colloidal nanoplatelets were about 66 meV.[11] We attribute more than two times higher values of splitting in our case to the influence of additional mode localization and, hence, a low mode volume in the designed tuneable microcavity, as well as a high photostability and low aggregation of the multishell QDs in the solution with high concentration. Moreover, in Fig. 2(a), it can be seen that the PL intensity of the LPB enhanced with an increase in the cavity mode energy, while the exciton fraction of the lower polariton was rising. It is well known that the emission from polaritonic states is proportional to the product of photonic fraction and population.[4] Thus, observed experimentally increase of the LPB emission with the decrease of the photonic fraction could only be due to a large increase of the LPB population. We consider two mechanisms of LPB population: photonic pumping from the emission of uncoupled states and vibrational scattering.[22,23] The first pathway of LPB population strongly depends on the overlap of between the spectra of the bare states emission and the energy of lower polariton.

The second one is proportional to the excitonic fraction of the LPB. The maximum emission intensity and the population of the LPB has been observed close to the energy of the emission of bare exciton states. This type of lower polariton intensity behaviour could be attributed to the significant impact of the photon pumping mechanism of the lower polariton population from the excitonic reservoir. Phonon scattering also should provide part of the polariton population for lower polariton states, which was highly detuned from the bare-state PL. However, the QDs had a large Stocks shift of about 110 meV, which resulted in a considerable overlap between the PL of the bare states and LPB for small detunings. In order to investigate the possibility of non-linear behaviour of the observed PL spectra, we measured the dependence of the PL spectra on the excitation power (Fig. 3).

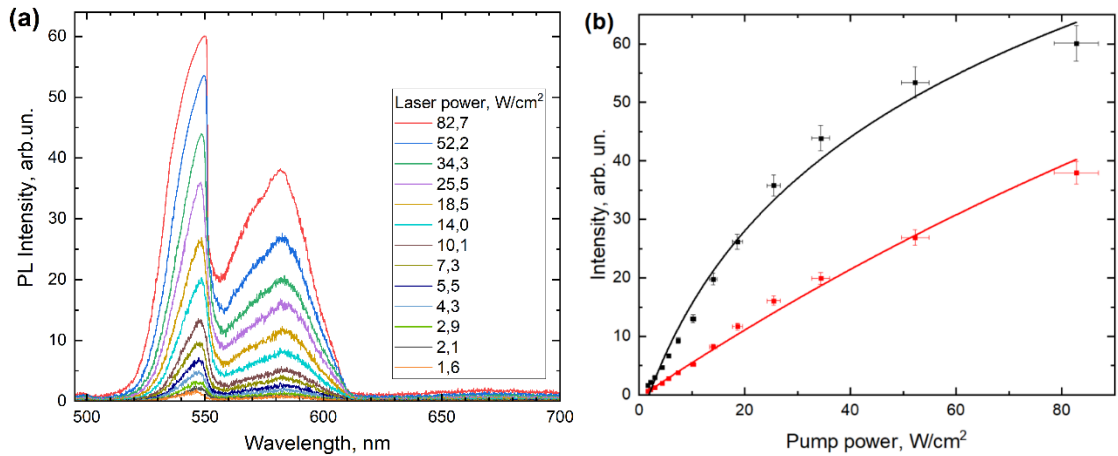

**FIG. 3.** The PL spectra (a) recorded at various non-resonant pump powers ($\lambda_{ex}$=450 nm). The extracted values of PL intensity (b) at the wavelengths corresponding to the emission from weakly (black) and strongly (red) coupled parts of the exciton transitions in the QDs ensemble.

Fig. 3(a) shows a rise in both emission maxima with an increase in the excitation power. Minor shifts of the PL maxima occurred at some excitation powers, which was due to the small random instability of the distance between the mirrors. The full width at half maximum was also preserved. We have extracted the intensities of PL maxima and plotted separately their dependence on the excitation power (Fig. 3(b)). As can be seen from Fig. 3(b), there no threshold behaviour was observed for LPB. For excitation powers below 40 W/cm$^2$, the emission from both states was found to be linear. For the pump energy above this value, we observed saturation of the emission for both weakly and strongly coupled states. We suggest that the observed behaviour is due to the absorption saturation of exciton transition and photodegradation of the part of the QDs ensemble.

To the best of our knowledge, this study is the first experimental demonstration of strong coupling between the photonic modes of a tuneable microcavity and localized excitons in colloidal semiconductor QDs. A coupling strength as large as 154 meV has been estimated from the fitting of the PL spectra measured at different cavity detunings with the use of the Jaynes–Cummings Hamiltonian. We observed emission from the weakly coupled part of the QD ensemble and LPB emission, while the UPB remained invisible. Analysis of the LPB population dependence on the cavity detuning demonstrated the optical pumping to be the main mechanism of LPB population in our experiments, which is due to the high quantum yields of the bare excitonic states and large Stocks shift. The latter provided sufficient overlap between the LPB and emission from the excitonic reservoir. We believe that, due to the high photostability, a quantum yield of up to 100%, wide absorption and narrow emission spectra, and room-temperature operation colloidal semiconductor QDs are promising for practical applications in polaritonics and polaritonic-based devices. Scalability of QD fabrication and the developed technology of coating them with surface ligands, which confers new chemical properties to QDs and allows targeted linking and control over the space position, pave the way towards the use of colloidal QDs in novel devices based on light–matter interaction.

This study was supported by the Ministry of Science and Higher Education of the Russian Federation (grant no. 14.Y26.31.0011). All developed procedures and approaches to the synthesis of nanomaterials presented in an article were supported by the Russian Science Foundation via the project no. 20-13-00358. I.N. acknowledges the support from the Ministry of Higher Education, Research and Innovation of the French Republic and the University of Reims Champagne-Ardenne. Y.R. acknowledges the support from the Basque Gouvernment (grant no. IT1164-19) and from the Spanish program MINECO (PID2019-111772RB-I00). We thank Vladimir Ushakov for the help with technical preparation of the manuscript.

**DATA AVAILABILITY**

The data that support the findings of this study are available from the corresponding author upon reasonable request.